\documentclass[11pt]{article}
\usepackage{times}
\usepackage{sidecap}
\usepackage{geometry}
\usepackage{floatrow}
\usepackage{float}
\usepackage{hyperref}
\usepackage{graphics,graphicx}
\usepackage{subcaption}
\usepackage{epsf,wrapfig}
\usepackage[export]{adjustbox}

\usepackage{natbib,bm}
\usepackage{paralist}
\let\olditem\item
\renewenvironment{thebibliography}[1]{%
  \section*{\refname}
  \let\par\relax
  \renewcommand{\item}[1][]{\olditem[\textbullet]}%
  \inparaenum}{\endinparaenum}

\newcounter{aff}

\DeclareGraphicsExtensions{.jpg,.pdf,.png,.eps}

\usepackage{caption}
\usepackage{xcolor}

\usepackage{tcolorbox}

\bibpunct{(}{)}{;}{a}{}{,}

\usepackage{soul}

\definecolor{darkcyan}{rgb}{0.0, 0.0, 0.7}

\definecolor{boxcolor}{RGB}{193, 229, 255}
\definecolor{figboxcolor}{RGB}{193, 229, 255}
\definecolor{boxcolorgray}{RGB}{227, 227, 227}
\definecolor{darkcyan}{rgb}{0.0, 0.0, 0.7}

\sethlcolor{boxcolorgray}

\def\gtrsim{\lower 2pt \hbox{$\, \buildrel {\scriptstyle >}\over
{\scriptstyle \sim}\,$}}
\def\lesssim{\lower 2pt \hbox{$\, \buildrel {\scriptstyle <}\over
{\scriptstyle \sim}\,$}}

\def\approxlt{\lower.2em\hbox{$\buildrel < \over \sim$}}
\def\approxgt{\lower.2em\hbox{$\buildrel > \over \sim$}}

\def\be{\begin{equation}}
\def\ee{\end{equation}}
\def\bea{\begin{eqnarray*}}
\def\eea{\end{eqnarray*}}

\def\jref@jnl#1{{\rm#1}}
\def\aj{\jref@jnl{AJ}}
\def\araa{\jref@jnl{ARA\&A}}
\def\apj{\jref@jnl{ApJ}}
\def\apjl{\jref@jnl{ApJ}}
\def\apjs{\jref@jnl{ApJS}}
\def\ao{\jref@jnl{Appl.~Opt.}}
\def\apss{\jref@jnl{Ap\&SS}}
\def\aap{\jref@jnl{A\&A}}
\def\aapr{\jref@jnl{A\&A~Rev.}}
\def\aaps{\jref@jnl{A\&AS}}
\def\azh{\jref@jnl{AZh}}		
\def\baas{\jref@jnl{BAAS}}
\def\icarus{\jref@jnl{Icarus}}
\def\jrasc{\jref@jnl{JRASC}}
\def\memras{\jref@jnl{MmRAS}}
\def\mnras{\jref@jnl{MNRAS}}
\def\pra{\jref@jnl{Phys.~Rev.~A}}
\def\prb{\jref@jnl{Phys.~Rev.~B}}
\def\prc{\jref@jnl{Phys.~Rev.~C}}
\def\prd{\jref@jnl{Phys.~Rev.~D}}
\def\pre{\jref@jnl{Phys.~Rev.~E}}
\def\prl{\jref@jnl{Phys.~Rev.~Lett.}}
\def\pasp{\jref@jnl{PASP}}
\def\pasj{\jref@jnl{PASJ}}
\def\qjras{\jref@jnl{QJRAS}}
\def\skytel{\jref@jnl{S\&T}}
\def\solphys{\jref@jnl{Sol.~Phys.}}
\def\sovast{\jref@jnl{Soviet~Ast.}}
\def\ssr{\jref@jnl{Space~Sci.~Rev.}}
\def\zap{\jref@jnl{ZAp}}
\def\nat{\jref@jnl{Nature}}
\def\iaucirc{\jref@jnl{IAU~Circ.}}
\def\aplett{\jref@jnl{Astrophys.~Lett.}}
\def\apspr{\jref@jnl{Astrophys.~Space~Phys.~Res.}}
\def\bain{\jref@jnl{Bull.~Astron.~Inst.~Netherlands}}
\def\fcp{\jref@jnl{Fund.~Cosmic~Phys.}}
\def\gca{\jref@jnl{Geochim.~Cosmochim.~Acta}}
\def\grl{\jref@jnl{Geophys.~Res.~Lett.}}
\def\jcp{\jref@jnl{J.~Chem.~Phys.}}
\def\jgr{\jref@jnl{J.~Geophys.~Res.}}
\def\jqsrt{\jref@jnl{J.~Quant.~Spec.~Radiat.~Transf.}}
\def\memsai{\jref@jnl{Mem.~Soc.~Astron.~Italiana}}
\def\nphysa{\jref@jnl{Nucl.~Phys.~A}}
\def\physrep{\jref@jnl{Phys.~Rep.}}
\def\physscr{\jref@jnl{Phys.~Scr}}
\def\planss{\jref@jnl{Planet.~Space~Sci.}}
\def\procspie{\jref@jnl{Proc.~SPIE}}

\def\ion#1#2{#1$\;${\small\rm\@Roman{#2}}\relax}

\newcount\lecurrentfam
\def\LaTeX{\lecurrentfam=\the\fam \leavevmode L\raise.42ex
\hbox{$\fam\lecurrentfam\scriptstyle\kern-.3em A$}\kern-.15em\TeX}
\def\sizrpt{
(\fontname\the\font): em=\the\fontdimen6\font, ex=\the\fontdimen5\font
\typeout{
(\fontname\the\font): em=\the\fontdimen6\font, ex=\the\fontdimen5\font
}}
\definecolor{dark}{rgb}{ 0.34, 0.09, 0.49}

\usepackage{fancyhdr,url}
\usepackage{transparent}
\usepackage{tablefootnote}
\usepackage{mathpazo}

\begin{document}

% Title and subtitle
\noindent
\begin{raggedright}
{\Huge\bfseries Planetary system architectures in the final stages of planet formation\\%[0.5em]
\smallskip 
%\Large 
\large \textit{How do debris discs trace planet assembly and volatile delivery?}
}\end{raggedright}\\

\medskip
\noindent
Patricia Luppe$^{1,*}$, Zoe Roumeliotis$^1$, Sebastian Marino$^2$, Quentin Kral$^3$, Joshua B. Lovell$^4$, Luca Matr\`a$^1$, Gianni Cataldi$^5$, Yinuo Han$^6$, Mark Booth$^7$, Isabel Rebollido$^8$, Camille Bergez-Casalou$^9$, Paul Huet$^3$, James Miley$^{10}$, Antranik A. Sefilian$^{11}$, Tim D. Pearce$^1$, \'Agnes K\'osp\'al$^{12}$, 
Stefano\\ Facchini$^{13}$, Korash Assani$^{14}$, Catherine Walsh$^{15}$, Danae Polychroni$^{16}$, Maria Drozdovskaya$^{14}$

\vspace{1em}
{
\footnotesize{
\noindent $^*$ luppep@tcd.ie; 
$^1$ School of Physics, Trinity College Dublin, the University of Dublin, College Green, Dublin 2, Ireland; 
$^{2}$ Department of Physics and Astronomy, University of Exeter, Stocker Road, Exeter EX4 4QL, UK; 
$^{3}$ LIRA, Observatoire de Paris, Université PSL, Sorbonne Université, Université Paris Cité, CY Cergy Paris Université, CNRS, 92190 Meudon, France; 
$^{4}$ Center for Astrophysics \textbar\, Harvard \& Smithsonian, 60 Garden Street, Cambridge, MA 02138, USA; 
$^{5}$ National Astronomical Observatory of Japan, Osawa 2-21-1, Mitaka, Tokyo 181-8588, Japan; 
$^{6}$ Division of Geological and Planetary Sciences, California Institute of Technology, 1200 E. California Blvd., Pasadena, CA 91125, USA; 
$^{7}$ UK Astronomy Technology Centre, Royal Observatory Edinburgh, Blackford Hill, Edinburgh EH9 3HJ, UK; 
$^{8}$ Centro de Astrobiolog\'ia (CAB) CSIC-INTA, Camino Bajo del Castillo s/n, 28692, Villanueva de la Cañada, Madrid, Spain; 
$^{9}$ Université Paris-Saclay, CNRS, Institut d’Astrophysique Spatiale, 91405 Orsay, France; 
$^{10}$ Joint ALMA Observatory, Alonso de Córdova 3107, Vitacura, Santiago, Chile; European Southern Observatory, Alonso de Córdova 3107, Vitacura, Santiago, Chile; Millennium Nucleus on Young Exoplanets and their Moons (YEMS), Chile; 
$^{11}$ Department of Astronomy and Steward Observatory, University of Arizona, Tucson, AZ 85721, USA; 
$^{12}$ Konkoly Observatory, HUN-REN RCAES, Budapest, Hungary; 
$^{13}$ Dipartimento di Fisica “Aldo Pontremoli”, Università degli Studi di Milano, via Celoria 16, Milano, 20133, Italy; 
$^{14}$ Department of Chemistry, Biochemistry and Pharmaceutical Sciences (DCBP), Universität Bern, Freiestrasse 3, 3012 Bern, Switzerland;
$^{15}$ School of Physics and Astronomy, University of Leeds, Leeds, LS2 9JT, UK;
$^{16}$ INAF - Osservatorio Astrofisico di Torino, Via Osservatorio 20, 10025, Pino Torinese (TO), Italy\\
}
}
%\vfill
%\vspace{-0.9cm}
%{
%\begin{flushleft}
%\textbf{Keywords:} \it planets,  planet-disc interactions, planet %formation, planetary systems, low-mass stars
%\end{flushleft}
%}
\vspace{-3mm}
\begin{tcolorbox}[boxsep=3pt,left=0pt,right=0pt,top=1pt,bottom=1pt,colframe=boxcolor,colback=boxcolor]
\justify
\textbf{Abstract:} 
%Limit of 1920 characters
The processes in the last stages of planet formation determine the final assembly and composition of planets.
%Systems form as protoplanetary discs: optically thick discs of gas and dust around newborn stars. 
After the protoplanetary discs dissipate, a system of planets and planetesimals -- analogues of the Asteroid and Kuiper Belts -- remains. Ongoing collisions between these bodies produce second-generation dust and gas that form a debris disc. %are of special importance in this framework. 
Resolved images of such belts allow us to trace the dynamical history of these systems, including giant impacts and planet-belt interactions, during the crucial era from 10 to a few 100 Myr of terrestrial planet formation. 
High-resolution mm and optical/near-IR imagers have revolutionised our view of the brightest planetesimal belts, providing first glimpses of diverse structures, many of which show plausible planet imprints, while ALMA has revealed gas in a growing number of systems. Yet these results are strongly biased toward the brightest, mostly young belts around early-type stars, and even for these, crucial details remain undetectable at the required resolution due to their typically low surface brightness.
A future (sub-)mm facility with an order of magnitude leap in continuum sensitivity and 10$\times$ higher line sensitivity 
%to complement the ELT's high-resolution capabilities, 
would transform our understanding of late-stage planet formation by enabling the first statistically meaningful surveys of faint outer planetesimal belts across all spectral types, imaging planetary collisions in the inner regions of planetary systems, and gas measurements that trace volatile delivery to terrestrial planets. For the first time, this would reach the faint majority of belts, rather than only the rare bright discs accessible today.  %This will for the first time extend down to
%A future ALMA requires a strong increase in sensitivity to image collisions, dust in the terrestrial planet zones, and to detect and resolve disc structures around all types of stars, down to the most common low-mass stars, where the majority of rocky planets are expected to mature in this latest stage of planet formation. 
\end{tcolorbox}
\vspace{0.7cm}
%\vspace{-0.8cm}
%\begin{center} %\transparent{0.4}
%\includegraphics[trim={10.30cm 0 10.3cm 0},clip,width=0.5\textwidth]{TWA7_title3.png} % Opacity

%\end{center}
%\vspace{2em}

\newpage{}
% Text

\vspace{-0.7cm}
\section{Scientific context and open questions}
% final phase of planet formation and debris discs
When dust- and gas-rich protoplanetary discs dissipate after a few Myr, they leave behind planets and planetesimal belts: rings of asteroidal and cometary bodies, analogous to the Solar System's Asteroid and Kuiper Belt \citep[e.g.][]{hughes-et-al-2018}. Continuing collisions between these planetesimals release second-generation dust and gas, observed as a \emph{debris disc}. This debris makes the belts detectable in scattered light and thermal emission across wavelengths ranging from the infrared to the millimetre.
We refer to cold outer belts as \emph{exo-Kuiper belts} (Sec.~\ref{sec:m_dwarfs}, \ref{sec:structures}) and to warm inner belts as \emph{exo-Asteroid belts} (Sec.~\ref{sec:terrestrial_planet_formation}, \ref{sec:structures}).
%a \emph{debris disc} consisting of asteroidal and cometary bodies (planetesimals). 
Planetary systems then enter the final phase of planet formation, lasting up to a few 100\,Myr \citep[e.g.][]{hughes-et-al-2018}, in which collisions, giant impacts, and volatile delivery continue to shape the final assembly of planets, while the planets interact with the surrounding disc \citep[e.g.][]{lisse-et-al-2009}. %The solid bodies in the disc continuously produce dust and gas. 
Dust and gas track these processes directly: their composition traces rocky and icy planetesimals, and thus the bulk composition of rocky planets.

 \vspace{-0.2cm}
\begin{figure}[ht]
    \centering
    \includegraphics[width=1.0\linewidth]{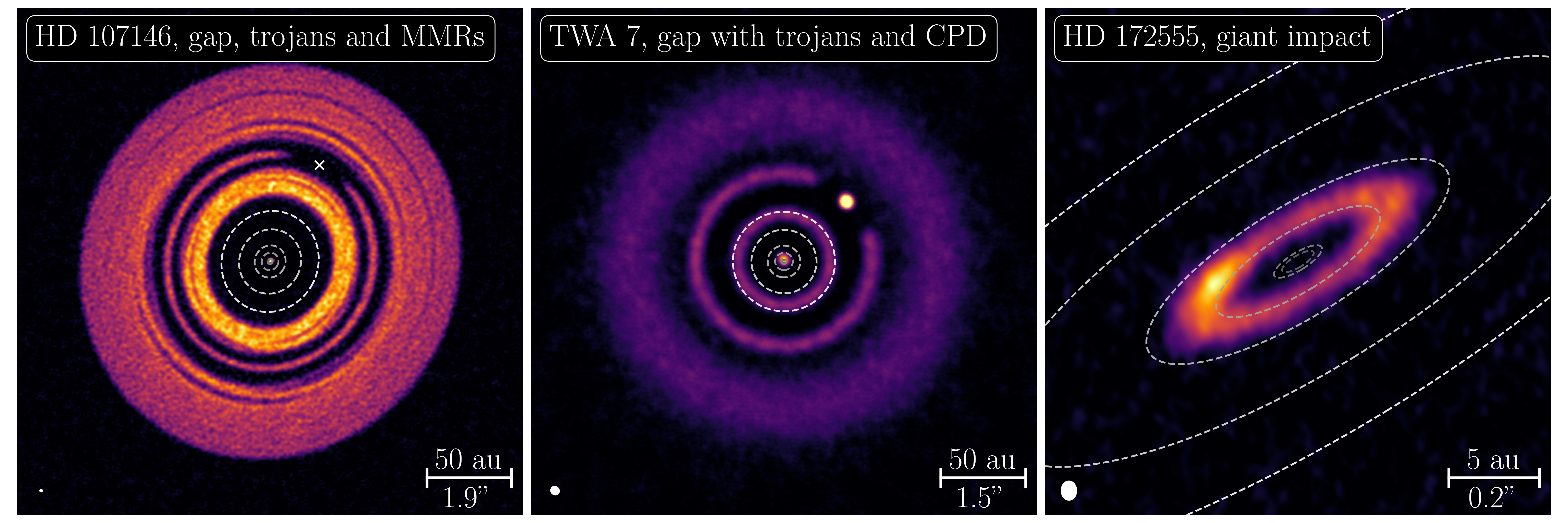}
    \vspace{-6mm}
    \caption{Simulations of 3 debris discs with predicted substructures for a future mm interferometer at 0.9~mm with %a 0.3~$\mu$Jy/beam sensitivity (512~GHz bandwidth and 255 antennas, 12~m). 
    the assumed ALMA2040 sensitvity (Table~\ref{tab:tech_requirements}). \textbf{Left:} G2V HD~107146 (${\sim}$100~Myr) with a putative Saturn-mass planet (represented by an 'x') embedded in the disc with co-orbital material. \textbf{Middle:} M2V TWA~7 (${\sim}$10~Myr) with a debris disc, a planet candidate surrounded by a circumplanetary disc and co-orbital material. \textbf{Right:} A7V HD~172555 (${\sim}$20~Myr) with an asymmetric post-impact debris disc in the $\sim$few au region. Dashed ellipses: Solar System planets' orbits from Earth to Neptune for reference. The beam size is shown in the bottom left of each panel.} 
    \label{fig:debris_disc_sim}
\end{figure}

% giant impacts
\vspace{-0.2cm}
Our view of both gas and dust in this phase is severely sensitivity-limited. Giant impacts between planetary embryos shape the final assembly of terrestrial planets and their moons \citep[e.g.][]{rhee-et-al-2008}, yet the dust and gas released within a few au of the host star has been resolved in only one system.
%However, these processes remain poorly constrained due to the lack of resolved observations of the dust and gas produced in these events within a few au from the host star. 
% M dwarfs
Bright exo-Kuiper belts are detected around ${\sim 20\%}$ of nearby AFGK stars \citep{thureau-et-al-2014, sibthorpe-et-al-2018}, but remain elusive around M~dwarfs, the most common stars and hosts of most known terrestrial planets \citep{mulders-et-al-2015}, whose belts are thought to be colder and fainter \citep[e.g.][]{lestrade-et-al-2025}. 
%As a result, only a handful of debris discs have been resolved around later-type stars due to sensitivity limitations.
%Structures
Planets interacting with debris discs drive collisions \citep[e.g.][]{mustill-wyatt-2009} and transport material from the outer disc to interior regions \citep[e.g.][]{bonsor-et-al-2014}, sculpting gaps, rings, clumps, asymmetries, and eccentricities \citep[e.g.][]{crotts-et-al-2024, lovell-et-al-2026}. These substructures can constrain the orbits, masses, and migration histories of otherwise undetectable Neptune- and Jupiter-analogues \citep[e.g.,][]{pearce-et-al-2022}.
%Structures old
%Interactions between planets and belts can drive collisions in the belts and transport material from outer belts to interior regions \citep[e.g.][]{marino-et-al-2018}, leaving observable imprints in the belts such as gaps, rings, clumps, asymmetries, and/or eccentricities \citep[e.g.][]{crotts-et-al-2024, lovell-et-al-2026}. 
%Depending on the exact substructures of a planetesimal belt, it is possible to place constraints on orbits, masses, and, in some cases, migration histories of otherwise undetectable ice or gas giant planets, analogous to Neptune or Jupiter \citep{pearce-et-al-2022}.
%Gas
Gas is expected in every debris disc, yet ALMA has detected it only in $\sim$20 systems, around young early-type stars \citep[e.g.][]{moor-et-al-2025}.
%A leap in line sensitivity is needed to characterise this component across host-star spectral types and ages. 
Measuring the gas abundance, distribution, and kinematics constrains planetesimal composition \citep[e.g.][]{matra-et-al-2017}, late-stage accretion of planetary atmospheres \citep{kral-et-al-2020}, and volatile delivery to terrestrial planets \citep{kral-et-al-2024}. 

In each case, the main limitation is sensitivity: a leap in continuum sensitivity would reveal the true debris disc population rather than the brightest $\sim$25\%, including belts around M dwarfs and individual giant impacts.  %belts around the most common stars in the Galaxy and image debris of individual giant impacts, while 
A leap in line sensitivity would characterise gas across spectral types and ages.
%Gas old
%ALMA detected gas in a small sample of bright debris discs hosted largely by young (tens of Myr) early-type stars \citep[e.g.][]{moor-et-al-2017}, despite being expected to be present in any debris disc at some level. A leap in line sensitivity is needed to systematically characterise the gaseous component of debris discs across host-star spectral types and ages. Measuring the gas abundance, distribution, and kinematics provides key information about the composition of planetesimals \citep[e.g.][]{matra-et-al-2017}, late-stage accretion of planetary atmospheres \citep{kral-et-al-2020} and the delivery of volatiles to terrestrial planets \citep{kral-et-al-2024}.
%of planets (e.g.\ after giant impacts) and planetesimals \citep[e.g.][]{matra-et-al-2017}. Planets interacting with the gas can be detected via their kinematic signatures. Finally, gas observations constrain late-stage accretion of planetary atmospheres \citep{kral-et-al-2020} and the delivery of volatiles (in particular water) to terrestrial planets \citep{kral-et-al-2024}.
A full picture of these final stages therefore requires a new (sub-)mm facility to address four key questions:
\vspace{-1mm}
%\begin{tcolorbox}[boxsep=3pt,left=0pt,right=0pt,top=0pt,bottom=0pt,colframe=boxcolor,colback=boxcolor]
\begin{enumerate}%[itemsep=0mm]
    \item How do terrestrial planets assemble through impacts in the latest stages of formation?
    \item What fraction of M~dwarfs, the most common planet hosts, harbour exo-Kuiper belts, and how are these icy reservoirs structured?
    %\item Do rocky planetary systems around M~dwarfs, the most common stars in the galaxy, host debris discs (ice reservoirs) in their outer regions?
    \item Do Neptune- and Jupiter-analogues shape debris discs, and how do these interactions affect planetary systems in the latest stages of formation?
    \item How does debris-disc gas depend on system age and spectral type, and how does it affect accretion and the volatile delivery to terrestrial planets?
    %How does the gaseous component of debris discs depend on system age and host star spectral type, and how is it affecting the late-stage accretion of planetary atmospheres and the delivery of volatiles to terrestrial planets?
\end{enumerate}
%\end{tcolorbox}
 %\vspace{-0.15cm}
 
\noindent
ALMA2040 could answer all four, 
complementing current and planned facilities (Sec.~\ref{sect:synergies}). We assume an ALMA2040 has 200-300 12~m antennas with 512~GHz bandwidth, reaching 0.3~$\mu$Jy/beam in 1~h at 0.9~mm (Table~\ref{tab:tech_requirements}).
Fig.~\ref{fig:debris_disc_sim} shows what ALMA2040 would reveal in three representative systems, spanning giant impacts (Sec.~\ref{sec:terrestrial_planet_formation}), M~dwarf belts (Sec.~\ref{sec:m_dwarfs}) and resolved substructures (Sec.~\ref{sec:structures}), while Sec.~\ref{sec:gas} quantifies the gains for gas.
%For example, the coronagraphs on the Extremely Large Telescope and Habitable Worlds Observatory will push optical/NIR capabilities tracing small dust grains. Together with Gaia's astrometric planet yield, these will reveal planets shaping discs. However, they leave a crucial gap; only a new (sub-)mm interferometer can image large (mm-sized) dust grains. Observations of both small and large dust grains are needed to elucidate processes impacting grain dynamics \citep[e.g.][]{olofsson-et-al-2022}. Thus, ALMA2040 would provide a crucial link for tracing disc structures and grain dynamics across all spatial and temperature regimes.

%Constraints on this outer exoplanet population will complement: JWST direct imaging, (E-ELT direct imaging?), and Roman ST population studies.

%\vspace{-6mm}

%\cc{Suggestions for content:}
%\cc{\begin{itemize}
%     \item what are debris discs
%     \item why are they important
%     \item what have we learnt already
%     \item what do we want to learn in future / why do we need ALMA2040 for it
%\end{itemize}}

%\newpage
% \vspace{-18pt}
%\begin{figure} %\vspace{-1cm}
%%\floatbox[{\capbeside\thisfloatsetup{capbesideposition={right,top},capbesidewidth=9.1cm}}]{figure}[\FBwidth]
%{\caption{\it Placeholder text and image, verbosely expanded to span the full width of the page. 
%\vspace{-0.7cm}\vspace{-0.5cm}}
%}
%{\includegraphics[width=7cm]{figure_cover.png}}
%\vspace{-0.5cm}
%\end{figure}
\vspace{-5mm}
\section{Science Case for the 2040's}
\vspace{-1mm}
\subsection{Terrestrial planet formation: the aftermath of planetary impacts}\label{sec:terrestrial_planet_formation}
 \vspace{-1mm}
 % NEW
 Terrestrial planet formation in the inner regions of planetary systems continues after the gas-rich protoplanetary disc has dissipated, with collisions between planetary embryos (giant impacts) dominating late-stage interactions and releasing dust and gas into the inner few au between $\sim$10 and a few 100 Myr \citep[e.g.][]{morbidelli-et-al-2012}. Evidence suggests an era of giant impacts in the Solar System, e.g., Earth's Moon, Mercury's iron enrichment, and the Martian hemispheric dichotomy \citep[][and references therein]{wyatt-and-jackson-2016}. Despite this being a crucial phase, it remains largely unconstrained in extrasolar systems because %it is difficult to resolve 
 the dust lies within a few au of the star. %with current facilities. 
 (Sub-)mm imaging is ideal for this as the disc-to-star contrast is orders of magnitude higher than at shorter wavelengths, revealing the terrestrial regions free of stellar contamination. To date, only HD~172555, at $\sim$29 pc -- one of the closest giant-impact systems -- has been resolved in both dust and CO gas emission with ALMA \citep{roumeliotis-et-al-2026}. The CO gas, at $\sim$5~au from the star, would be quickly photodissociated if released by a collisional cascade, so it points to planetary atmospheric stripping during the impact.
 % As the CO gas is only $\sim$5 au from the central star, it is expected to be the result of atmospheric stripping during the impact. %Despite the deep integration, the ALMA data lacked the sensitivity to resolve the dust's inner edge at $<$2.3 au. 
\par Tens of other $\sim$10 to $\sim$100 Myr-old systems show warm IR dust (100s of K) in the terrestrial region that is too bright to be explained by Asteroid belts alone. These \textit{extreme debris discs} \citep[e.g.][]{wyatt-et-al-2007, balog-et-al-2009, fujiwara-et-al-2012} are thought to result from large collisions between Moon- to Mars-sized bodies and can vary on timescales of months to years \citep[e.g.][]{moor-et-al-2021, su-et-al-2026}. Characterising them requires the sensitivity to survey large samples and the resolution to image gas and dust structures in their terrestrial regions.

\par
\begin{figure}[ht]
 \begin{subfigure}{.5\textwidth}
      %\centering
     %\includegraphics[width=\linewidth, left]{ALMA2040_dustmassdetectability.png}
     \includegraphics[width=\linewidth, trim={1mm 0mm 0mm 3mm}, clip]{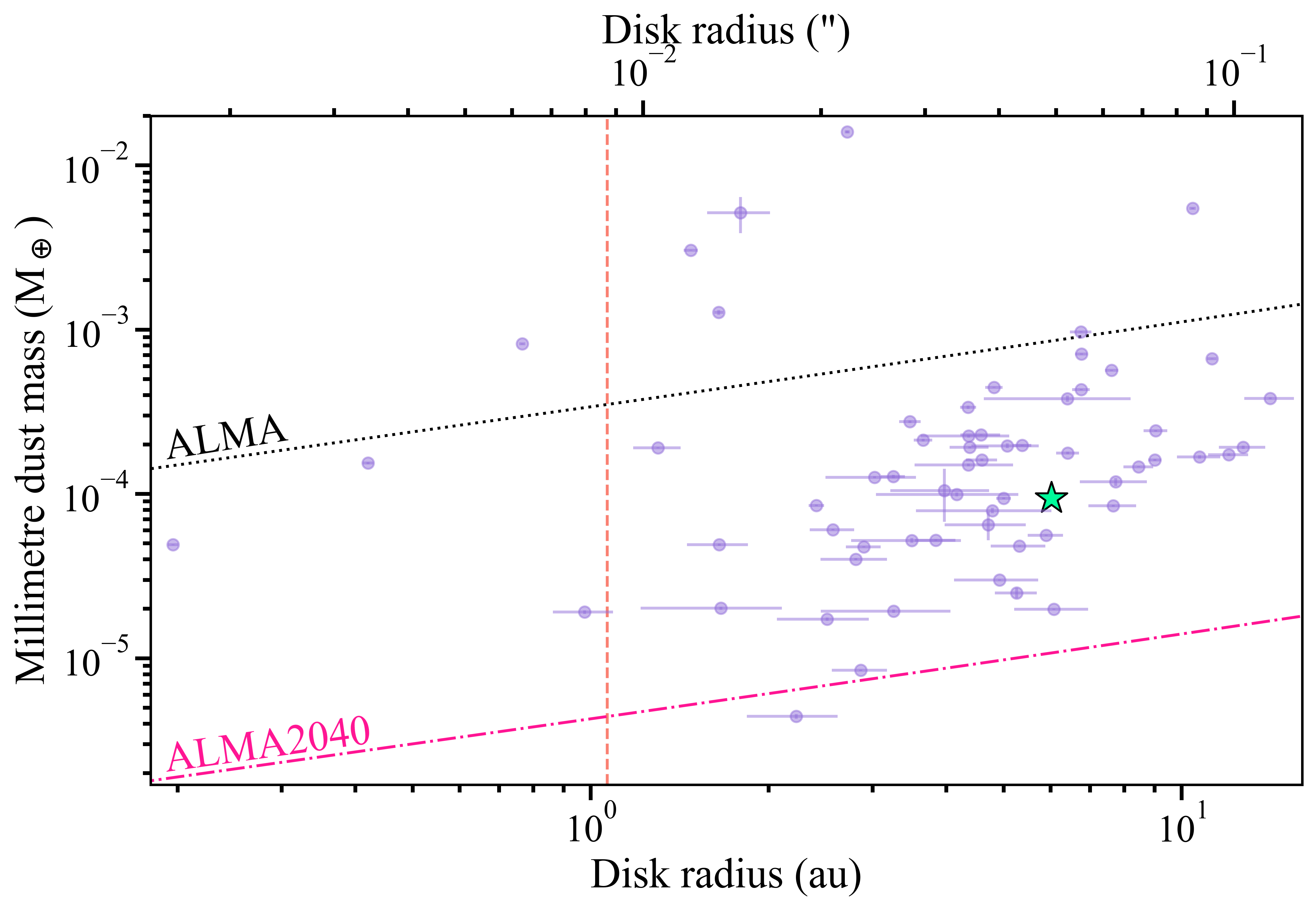}
     %trim={left bottom right top}
     \vspace{-4mm}
 \end{subfigure}\hspace{-2.4mm}
 \begin{subfigure}{.251\textwidth}
     %\centering
      \raisebox{9mm}[\height][\depth]{
      \includegraphics[width=\linewidth, trim={1mm 1mm 1mm 1mm}, clip]{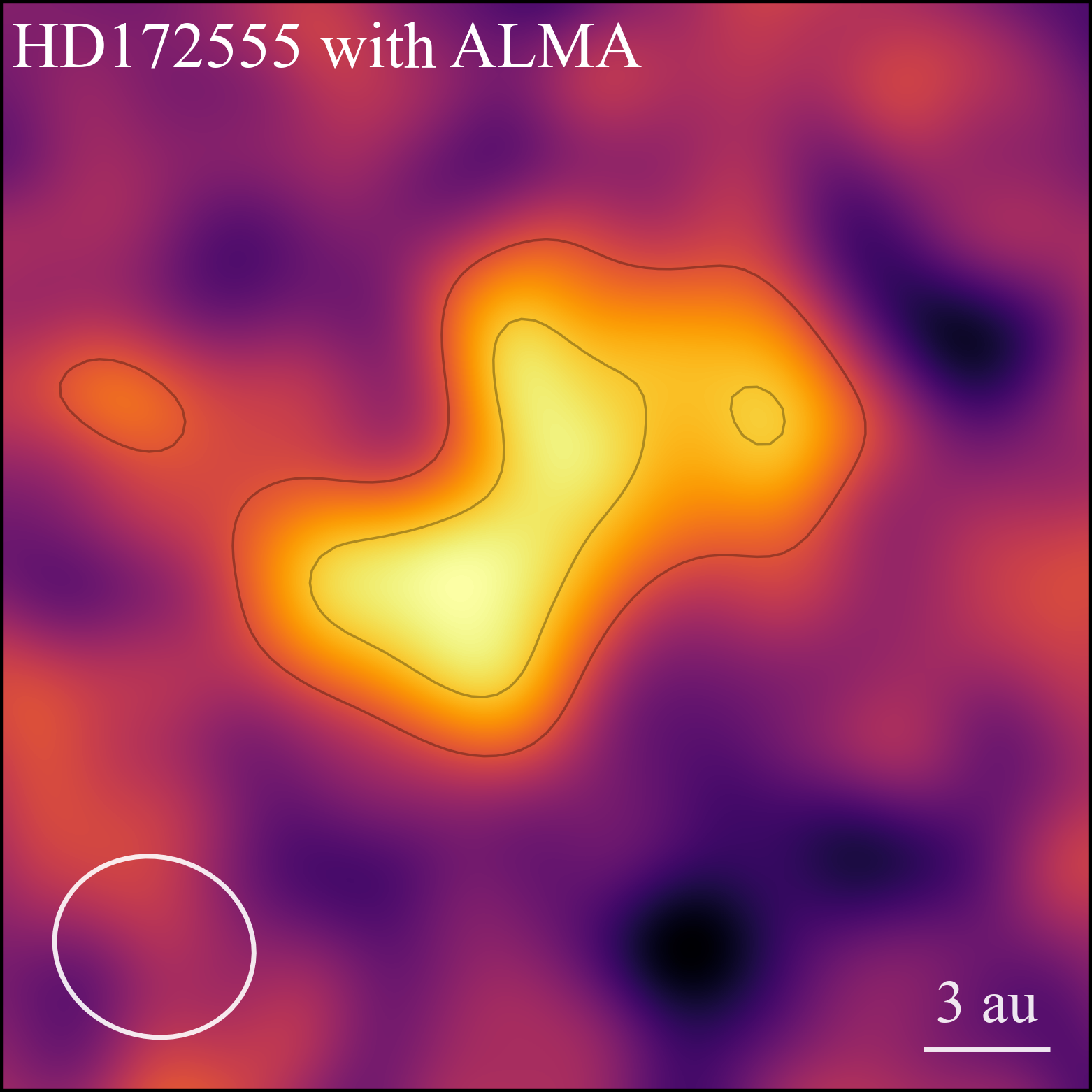}
      }
      %trim={left bottom right top}
 \end{subfigure} \hspace{-1.5mm}
 \begin{subfigure}{.251\textwidth}
     %\centering
      \raisebox{9mm}[\height][\depth]{
      \includegraphics[width=\linewidth, trim={1mm 1mm 1mm 1mm}, clip]{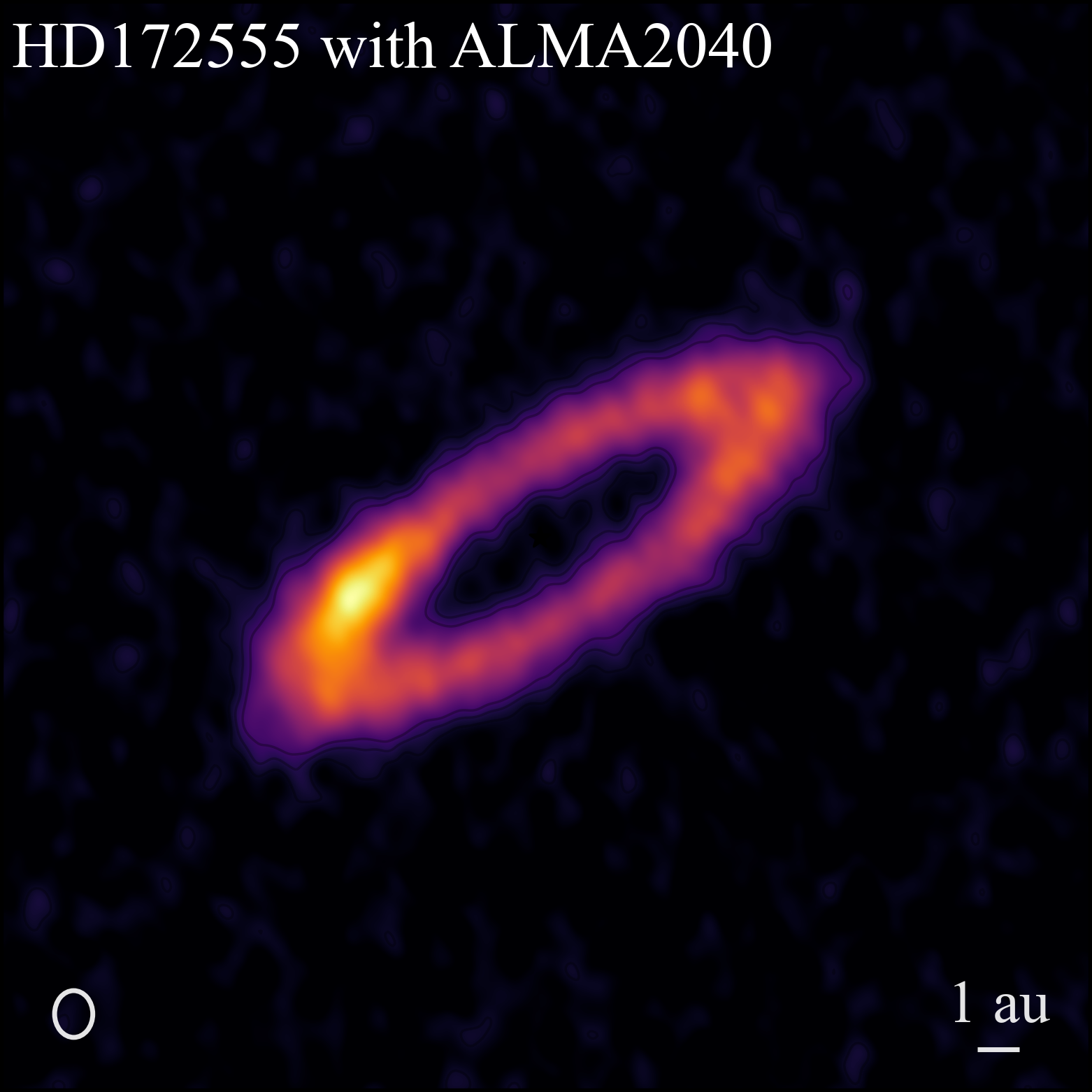}
      }
      %trim={left bottom right top}
 \end{subfigure}
 \vspace{-3.5mm}
\caption{Left: mm dust masses versus disc radius for giant impact candidate systems. The black dotted line is ALMA's current continuum sensitivity, and the pink dash-dotted line is ALMA2040's (Table~\ref{tab:tech_requirements}), both for 1~h at 0.9~mm, plotted for the median distance and disc temperature of the sample. The vertical dashed line marks the required ALMA2040 resolution, which corresponds to $\sim$1 au (median distance). The green star marks HD~172555. Middle: current ALMA data for HD 172555 \citep{roumeliotis-et-al-2026}. Right: Simulated ALMA2040 data for HD~172555.}
 \label{fig:giant impact systems}
 \end{figure}

 Fig. \ref{fig:giant impact systems}, left, shows the predicted mm-dust mass as a function of disc radius for a \textit{Spitzer} Infrared Spectrograph sample of systems with extreme levels of dust in their terrestrial regions \citep{chen-et-al-2014}. With one hour of integration time per target, ALMA only detects a handful of giant impact candidates and covering the full sample would need $>$6 years. The faintest target alone would need 528 days of integration time, whereas ALMA2040 would detect it in 2 hours and complete the entire survey in 8.8 hours of integration, with 95\% of impact candidates detectable within a single hour. %In contrast, that same target would be detectable in just 8.25 hours by ALMA2040, and the entire survey could be done in 35 hours. With the increased continuum sensitivity of ALMA2040 (Sec.~\ref{sec:TRs}), 89\% of giant impact candidate systems would be detectable with one hour of integration time with ALMA2040. 
Such a survey with ALMA2040 would measure the mm-dust masses of these systems and constrain the occurrence of warm dust above 10$^{-6}$ M$_\oplus$ within the median distance of 120 pc, four orders of magnitude below the debris released in the Moon-forming impact \citep{canup-2004}. 

\par %To characterise and better understand giant impact systems, 
Resolved images are needed to characterise these systems. %With current facilities, this can only be done for one system, 
Even for HD 172555, 16 hours of on-source time with current ALMA were not enough to resolve the dust’s inner edge at $<$2.3~au in the terrestrial planet-forming region (Fig. \ref{fig:giant impact systems}, middle). Resolving the inner edge of the dust disc is crucial for determining the width of the disc, which is directly related to the mass of the largest impacting planet \citep[e.g.][]{jackson-et-al-2014, schneiderman-et-al-2021}. JWST cannot do this due to its limited spatial resolution and contrast at close separations. In the 2040s, HWO and the ELT will image small grains in these systems (Sec.~\ref{sect:synergies}), but only a (sub-)mm facility traces the large grains that remain close to the radial location of the impact. Resolved images would also reveal dust-structure asymmetries, which indicate the location of the impact and the amount of time that has passed since it occurred \citep{jackson-et-al-2014}. With ALMA2040, many giant impact systems could be resolved and monitored (Fig.~\ref{fig:giant impact systems}), enabling large surveys to study the occurrence of giant impacts and routine high-resolution imaging to characterise the composition and evolution of debris and thus the building blocks of terrestrial planets.
% enabling observational constraints on this important era of terrestrial planet formation.  

\vspace{-4mm}

 % OLD
% Terrestrial planet formation continues after the gas-rich protoplanetary disc has dissipated, with collisions between forming planetesimals (giant impacts) dominating late-stage interactions \citep[e.g.][]{morbidelli-et-al-2012}. %(e.g. %Chambers 2001;Morbidelli et al. 2012).
% Giant impacts release dust and gas into the inner few au regions. Despite this being a crucial phase of planet and moon formation, it remains observationally unconstrained because it is difficult to resolve the dust in the terrestrial zones (up to a few au) of planetary systems with current facilities. Only one post-giant impact disc has resolved dust and CO gas: HD 172555, one of the closest giant impact systems \citep{roumeliotis-et-al-2026}. Despite the deep integration, the ALMA data lacked the %surface brightness 
% sensitivity to resolve the dust's inner edge at $<$2.3 au%from the central star
% . There are other $\sim$10-100 Myr-old systems with warm IR dust (100s of K) in the terrestrial region that are too bright to be explained by asteroid belts, and could result from giant impacts \citep[e.g.][]{wyatt-et-al-2007, fujiwara-et-al-2012}, but these systems are too faint and/or compact to be resolved by current facilities. %A (sub-)millimetre facility with improved sensitivity and resolution is vital for a better understanding of the era of giant impacts, as well as for constraining the post-impact evolution of dust and gas. 

%%%%%%%%%%%%%%%%%%%%%%%%%%%%%%%%%%%%%%%%%%%%%%%%%%
%-------- M dwarfs / disc populations
%%%%%%%%%%%%%%%%%%%%%%%%%%%%%%%%%%%%%%%%%%%%%%%%%%
 \vspace{-7mm}
\subsection{Detecting and resolving debris discs around cool M dwarf stars}\label{sec:m_dwarfs}
\begin{figure}[ht]
\begin{centering} 
%\hspace{1.5cm}
\includegraphics[width=\textwidth]{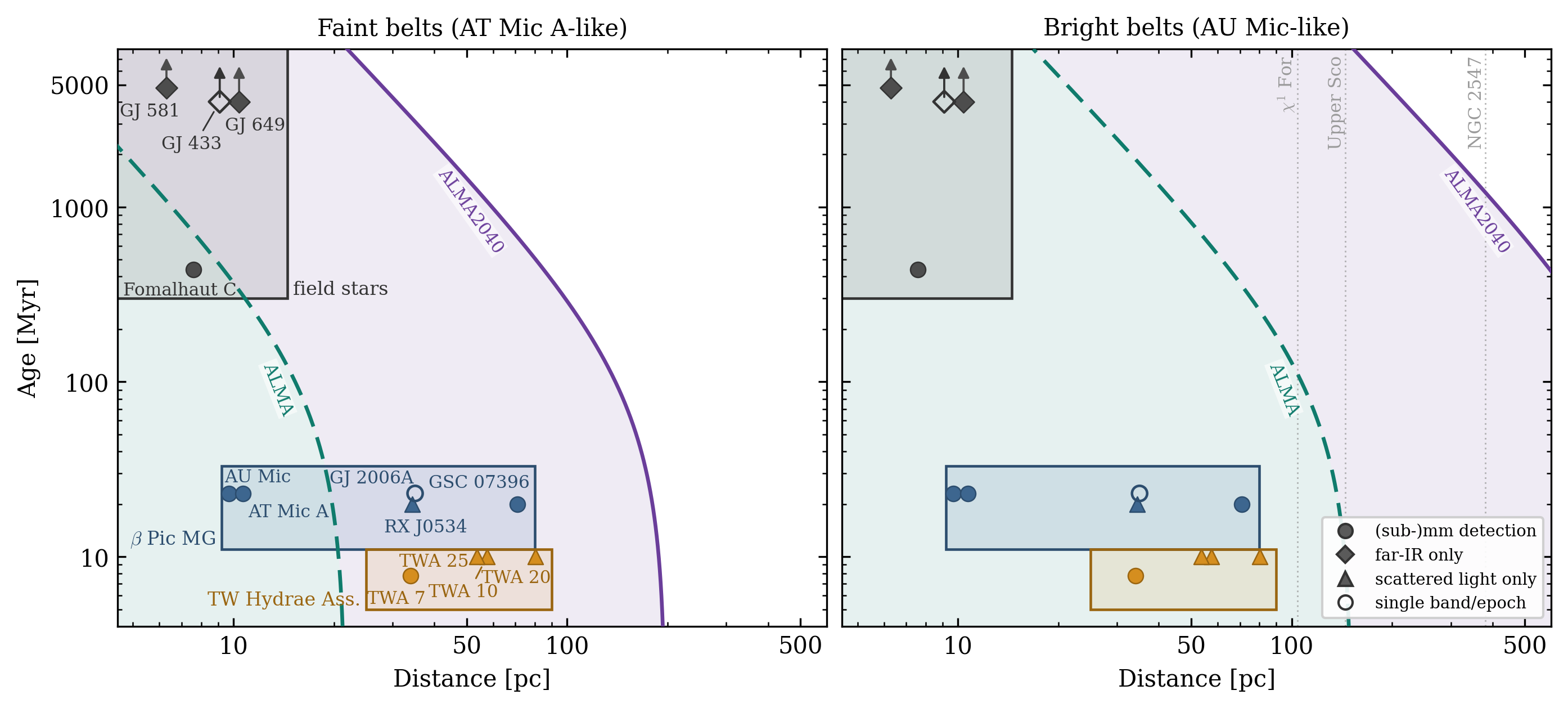}
\vspace{-6mm}
\caption{Known M~dwarf debris discs in the age-distance plane. %, with detection ranges of ALMA (teal) and ALMA2040 (purple) for a compact configuration. 
Curves show maximum distances at which ALMA (dashed teal) and ALMA2040 (solid purple) detect a belt as faint as the faintest (AT MiC A; left) or as the brightest (AU Mic; right) confirmed M dwarf belt, at 3$\sigma$ in 1~h per pointing at 0.9~mm. Teal shading marks the range of current ALMA, purple the range newly reached by ALMA2040.
Marker shapes show the detection method: (sub-)mm (circles), far-infrared only (diamonds), scattered light only (triangles). Filled symbols are robust detections, and open symbols mark unresolved single band/epoch excesses. Boxes show the ranges of discs around field stars (black), BPMG (blue), TWA (gold). Dotted lines (right) mark the median distance of young, distant clusters.
}
\vspace{-2mm}
\label{fig:Mdwarf_dist_vs_age}
\end{centering} 
\end{figure}

 \vspace{-1mm}
M~dwarfs make up $\sim$70\% of stars \citep{bochanski-et-al-2010} and are the most favourable hosts for detecting terrestrial planets \citep{shields-et-al-2016}, yet their exo-Kuiper belts remain almost unexplored.
%Four decades of infrared and optical all-sky searches from \emph{IRAS}, \emph{WISE}, to \emph{Gaia}, together with many targeted surveys 
In decades of infrared and optical searches \citep[e.g.][]{gautier-et-al-2007, lestrade-et-al-2009, avenhaus-et-al-2012, sgro-song-2021}, %only brought up one confirmed M~dwarf debris disc: AU~Mic. Most other candidates turned out to be stellar flares, background galaxies or were reclassified to protoplanetary discs.
most candidates proved to be stellar flares, background galaxies or protoplanetary discs.
%Targeted surveys \citep[e.g.][]{gautier-et-al-2007, lestrade-et-al-2009, avenhaus-et-al-2012, sgro-song-2021}, only around a dozen planetesimal belts have been detected - ten of them resolved or independently detected in two bands or epochs - and even fewer have been resolved. 
\emph{Herschel} was the first to reach the cold dust around the nearest field M~dwarfs, with detection rates of $\sim$2-5\% 
\citep[black in Fig.~\ref{fig:Mdwarf_dist_vs_age};][]{lestrade-et-al-2012, kennedy-et-al-2014, kennedy-et-al-2018}.
ALMA increased this to 11\% in the younger $\beta$ Pic moving group \citep[BPMG; blue in Fig.~\ref{fig:Mdwarf_dist_vs_age};][]{cronin-coltsmann-et-al-2023} and \emph{JWST} has since added detections in the TW Hydrae Association \citep[TWA; gold in Fig.~\ref{fig:Mdwarf_dist_vs_age};][]{palatnick-et-al-2025, crotts-et-al-2025} and BPMG \citep{ferrer-chavez-et-al-2026}. %, and confirmed TWA~25, but in scattered light, which traces micron-sized grains rather than the cold dust which traces the location of the planetesimals. %Only six have been detected in the millimetre.
Of the $\sim$13 known M~dwarf debris discs, %10 are robust (spatially resolved or independently detected in two bands or epochs), 
only 6 have been detected in the millimetre, and of these only four are well resolved. 
% 6 mm discs: Fom C, AU Mic, AT Mic A, GJ 2006, GSC, TWA 7 
% resolved in the mm: Fom C, AU Mic, TWA 7, GSC (marginally), GJ581 (marg.)
Yet a statistical analysis of \emph{Planck} data suggests that $\sim$10\% of M~dwarfs within 80~pc have debris emission \citep{nibauer-et-al-2020}, consistent with \citet{luppe-et-al-2020}, 
who argue that these belts are not absent, but simply too faint for current mm facilities. 

%This is because with their low surface brightness,  resolving them sufficiently well to characterise their structure requires long observing times. 
Current ALMA is well suited for bright belts around luminous stars \citep{matra-et-al-2025}, but reaching M~dwarf  belts requires a transformative sensitivity increase.
Fig.~\ref{fig:Mdwarf_dist_vs_age} shows the maximum distance to which an exo-Kuiper belt is detectable at 3$\sigma$ in 1 h per pointing at 0.9~mm with current ALMA (dashed teal) and ALMA2040 (solid purple), for the faintest (AT Mic A; left) and the brightest (AU Mic; right) confirmed M~dwarf belt, faded with age using the collisional evolution model of \citet{wyatt-2008}. The curves bend at the collisional timescale (here $t_{\rm c}=100$~Myr), beyond which dust mass and therefore flux falls as $1/t$.

\vspace{-2mm}

%Faint AT~Mic analogue discs can be detected when they are located closer than the thick lines (dashed teal for ALMA, solid purple for ALMA2040) and bright AU~Mic analogues can be detected up to the distance shown by the shaded area. The detection curves reach further distances for younger stars, because older stars are expected to have fainter discs. %The curves bend at the collisional timescale (here $t_{\rm c}=100$~Myr). Younger belts have barely ground down, and their brightness is nearly constant, while for older belts the dust mass and therefore the flux falls as $1/t$.
\vspace{-2mm}
All confirmed mm detections lie within current ALMA's reach. At 10~Myr, ALMA detects faint, AT~Mic-A-like discs out to $\sim$21~pc and ALMA2040 out to $\sim$190~pc (Fig.~\ref{fig:Mdwarf_dist_vs_age}, left), beyond ALMA's current reach for even the brightest, AU Mic-like belts. ALMA2040 would detect the youngest and brightest exo-Kuiper belts out to $\sim$1.3~kpc (Fig.~\ref{fig:Mdwarf_dist_vs_age}, right), bringing distant young clusters within range for the first time.
%reaching distant clusters.

%Without this, we do not know if planetesimal belts around M dwarfs are just faint or fundamentally different \citep{luppe-et-al-2020}, which might indicate differences in the formation and evolution of planets and planetesimals compared to hotter-type stars.

Sensitivity is not the only requirement: the nearest belts are also the largest on the sky. GJ~581 spans 8-19$''$ at $\sim$6~pc \citep{lestrade-et-al-2012}, AU~Mic $\sim$8$''$ \citep{macgregor-et-al-2013} and Fomalhaut~C $\sim$7$''$ \citep{cronin-coltsmann-et-al-2021}. Characterising the closest belts therefore requires ALMA2040 to combine its sensitivity gain with short spacings that recover scales of 10-20$''$ (Table~\ref{tab:tech_requirements}).

Detection is not characterisation. 
TWA~7, which hosts a planet \citep{lagrange-et-al-2025}, shows three rings in scattered light \citep{ren-et-al-2021, crotts-et-al-2026}, but is only marginally resolved by current ALMA \citep{bayo-et-al-2019}. Scattered light traces small grains that can be moved away from their parent belt by stellar winds or radiation forces. Only (sub-)mm imaging locates grains tracing the planetesimal positions. Fig.~\ref{fig:debris_disc_sim} (centre) shows TWA 7's belt observed with ALMA2040. With 1000s of nearby M-dwarfs, such a sensitivity boost would transform our view of planetary systems around the most common stars in the Galaxy.\\

%%%%%%%%%%%%%%%%%%%%%%%%%%%%%%%%%%%%%%%%%%%%%%%%%%
%-------- resolving substructures
%%%%%%%%%%%%%%%%%%%%%%%%%%%%%%%%%%%%%%%%%%%%%%%%%%
 \vspace{-8mm}
\subsection{Detailed debris disc substructures and planet-disc interactions}\label{sec:structures}
 \vspace{-0.1cm}
 %new -- JBL
Debris discs trace the formation and dynamical history of planetary systems. Just as Jupiter's and Neptune’s current orbits and migration histories sculpted the structures of the Asteroid and Kuiper belt respectively \citep[trapping planetesimals in mean motion resonances, clearing unstable orbital radii, and pumping up orbital inclinations and eccentricities;][]{Wisdom_1982, FernandezIp_1984, Malhotra_1995} the detailed spatial distribution of dust in debris discs can reveal the presence, orbits and masses of nearby, sculpting exoplanets \citep[see e.g.][]{Mouillet+1997, Wyatt+1999, Quillen_2006, pearce-et-al-2022}. Because many planets predicted from debris disc architectures lie well below the detection limits of current exoplanet-hunting facilities, debris disc substructures can offer unique views of the otherwise inaccessible population of exoplanets in the Galactic neighbourhood (typically cool Saturn- and Neptune-analogues).

\vspace{-4mm}
\begin{figure}[ht]
    %\centering
    %\begin{subfigure}[b]{0.495\textwidth}
    %\centering
    \includegraphics[width=0.495\linewidth]{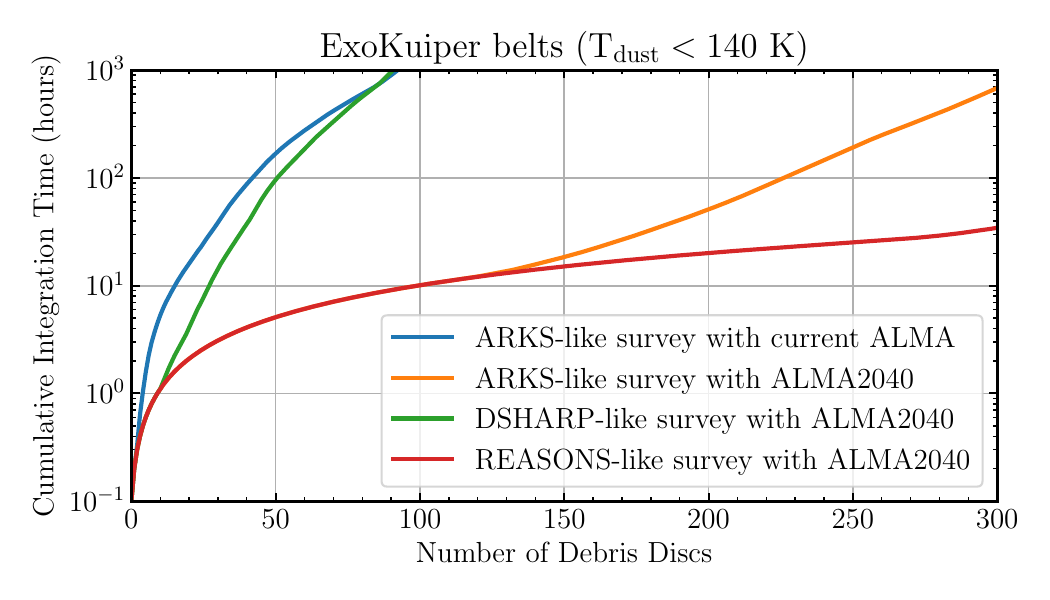}
    %\end{subfigure}
    %\begin{subfigure}[b]{0.495\textwidth}
    \includegraphics[width=0.495\linewidth]{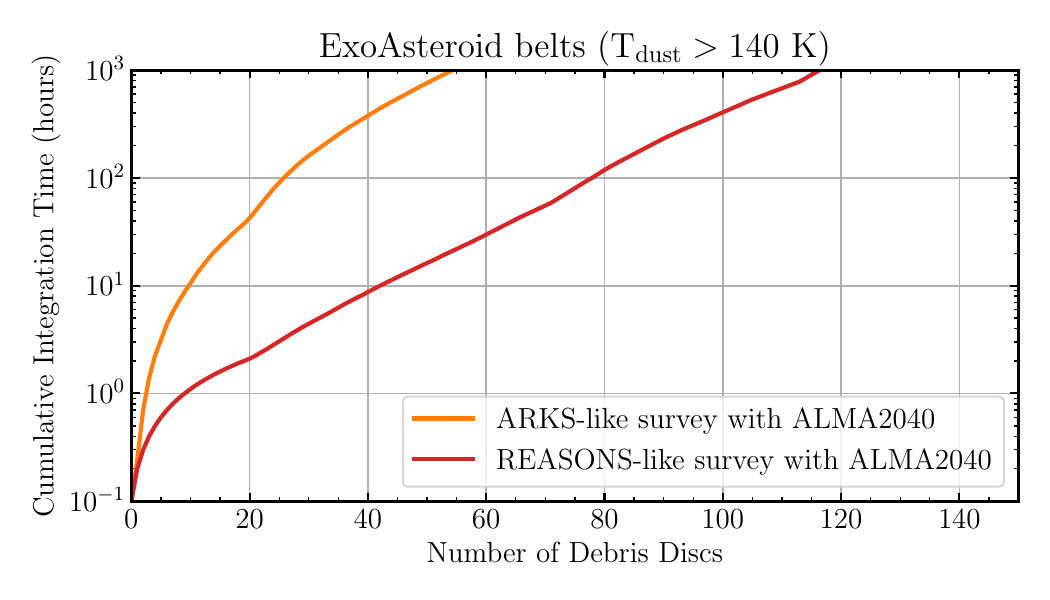}
    %\end{subfigure}
 
    \vspace{-4.5mm}
    \caption{Cumulative integration time curves for a range of surveys of cold exo-Kuiper belts (left) and warm exo-Asteroid belts (right), possible with current ALMA and those unlocked by ALMA2040 that will enable population studies of debris discs. Each curve corresponds to a survey with a resolution (as \% of the disc width) and sensitivity (as S/N in the image or after azimuthally averaging): ARKS-like \citep[azimuthally averaged S/N${\approx}10$ and beam=20\%,][]{marino-et-al-2026}, REASONS-like \citep[azimuthally averaged S/N${\approx}10$ and beam=100\%,][]{matra-et-al-2025}, and DSHARP-like \citep[image S/N${\approx}30$ and beam=10\%,][]{Andrews+2018}.} 
    \label{fig:time_survey_debris}
\end{figure}
\vspace{-2mm}
ALMA imaging of the 3D dust and gas structures of cold debris discs at tens of au has revealed diverse morphologies and substructures in the radial, azimuthal and vertical dimensions linked to the presence of planets and planetary system dynamical histories \citep[e.g.][]{dent-et-al-2014, matra-et-al-2019, Lovell+2025, Han+2026, Zawadzki+2026, MacManamon+2026}. 
Yet, these detailed studies are limited to a few cold, bright debris disc systems where sharp angular resolutions (${\sim}$10\% of disc radii) can be reached, and even then, only with very large time investments \citep[e.g. the ALMA Large Program ARKS required ${\approx}$200 hours of on-source time to study 24 systems, see][]{marino-et-al-2026}. 
Even for the brightest discs, ALMA’s current S/N is only moderate when averaging over multiple beams (with S/N ranging from a few to a few tens), which can lead to ambiguous architectural interpretations \citep{Han+2026}. Moreover, since the morphologies of these bright discs are likely unrepresentative of the wider debris disc population, what remains critically missing are high-resolution studies of fainter belt populations, with architectures and masses more akin to the Solar System’s cold Kuiper belt and inner warm Asteroid belt. 

With a transformative increase in sensitivity at (sub-)mm wavelengths (Sec.~\ref{sec:TRs}), ALMA2040 offers three capability enhancements to study debris disc substructures and planet-disc interactions. 
i) Detecting or ruling out features predicted by disc-planet interaction models; e.g. co-orbital populations~ (Fig.~\ref{fig:debris_disc_sim}), gaps, spiral arms, resonant populations and warps, %to distinguish between competing models through the distinct spatial signatures they predict, and convert currently 
turning ambiguous %or statistically-limited 
detections into robust constraints on unseen outer planet populations. 
ii) An order-of-magnitude increase in the population of discs available for high angular-resolution mm-studies: in 100 h, an ARKS-like survey would reach $>200$ exo-Kuiper belts with ALMA2040, compared with $<$40 belts with current ALMA(Fig.~\ref{fig:time_survey_debris}, left), shifting the field from individual case studies to population-level statistics across stellar host mass, age, and system architecture. 
iii) The first high-resolution views on inner exo-Asteroid belts: in 100 h, a REASONS-like survey would reach $>70$ exo-Asteroid belts with ALMA2040, which is not currently possible with ALMA (Fig.~\ref{fig:time_survey_debris}, right).
Exo-Asteroid belts steadily grind down, unlike the transient giant-impact debris of Sec.~\ref{sec:terrestrial_planet_formation}, and so open up a comparative window onto the processes sculpting the inner terrestrial planet-forming regions of planetary systems.
These goals require ALMA2040's continuum sensitivity, a modest ${\approx}2{\times}$ improvement in angular resolution, and short spacings to cover the nearest discs (Table~\ref{tab:tech_requirements}).

\vspace{-3mm}
\subsection{Gas as a tracer of the composition, late accretion, and evolution of planetary systems}\label{sec:gas}
\vspace{-1mm}
Debris discs were traditionally considered gas-poor, yet the past decade has revealed substantial amounts of gas in a growing number of these systems \citep{moor-et-al-2017, cataldi-et-al-2023}. Unlike the primordial, H$_2$-dominated gas in protoplanetary discs, this gas is generally thought to be secondary \citep[e.g.][]{smith-et-al-2026}, released by destruction, collision, or sublimation of volatile-rich planetesimals \citep{kral-et-al-2017, bonsor-et-al-2023}. %, except for the most massive discs where some ambiguity still exists \citep[e.g.][]{nakatani-et-al-2021}, . 
Its composition, spatial distribution, and kinematics therefore offer a view of late-stage planet formation that dust alone cannot provide. Yet, observations remain largely restricted to CO and to bright debris discs around A stars \citep{MacManamon+2026}. Nearly all are younger than $\sim$40~Myr, with the 440~Myr old Fomalhaut belt the only detection at a greater age \citep{matra-et-al-2017}, preventing systematic characterisation of the origin, prevalence, and impact of this gas on planets \citep[e.g.][]{kral-et-al-2020}.

\begin{figure}[ht]
    \centering
    \includegraphics[width=0.8\linewidth]{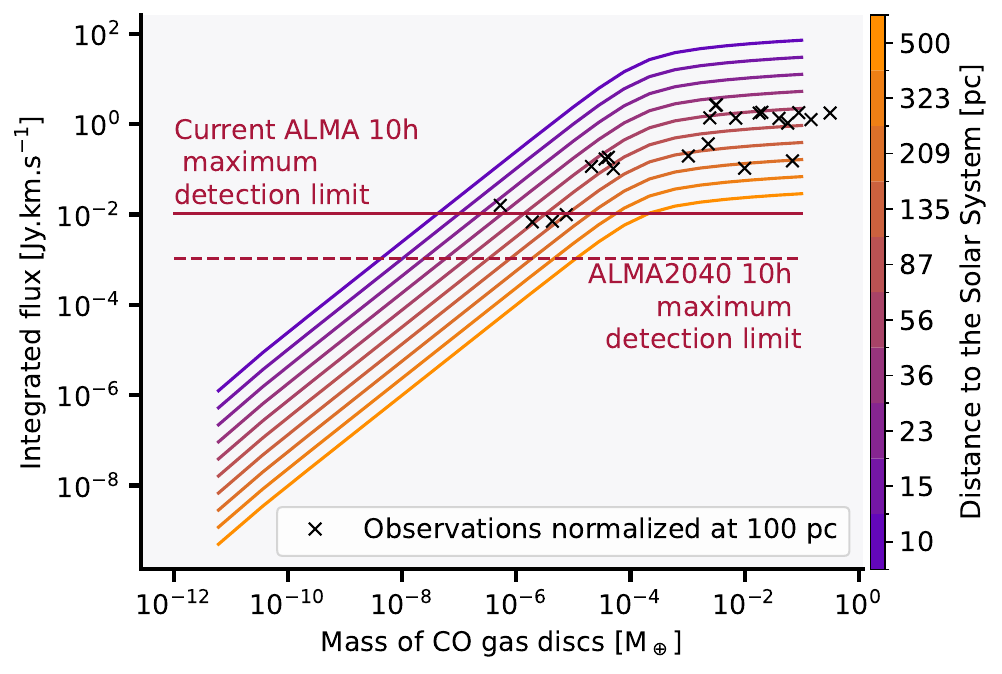}
    \vspace{-4mm}
    \caption{Integrated flux versus CO gas mass calculated for discs observed at different distances (colourbar). Black crosses show previous ALMA gas detections, with fluxes normalised to a distance of 100 pc. The ALMA2040 10~h detection threshold (dashed line) lies an order of magnitude below the current threshold (solid line), enabling the detection of discs with current CO gas masses out to 500~pc, or of $\sim$10$\times$ lower-mass discs in nearby systems at a given distance.} 
    \label{fig:fluxgas}
\end{figure}

ALMA2040 will transform this picture by resolving multiple molecular and atomic tracers both spatially and spectrally: %by enabling sensitive, spatially and spectrally resolved observations of multiple molecular and atomic tracers, including 
CO, C I and potentially H$_2$O and OH, CN and HCN, which have so far remained undetected in debris discs \citep[e.g.][]{matra-et-al-2018a}. 
%It will also reach gas in belts an order of magnitude older, and correspondingly less massive belts, than those accessible today, extending gas studies beyond A-type stars (see Fig.~\ref{fig:fluxgas}). 
It will detect gas masses an order of magnitude lower than today's limits at a given distance, or most discs with current gas masses out to $\sim$500~pc (Fig.~\ref{fig:fluxgas}), extending gas studies to older, less massive belts.
Measuring several species and isotopologues simultaneously will test whether the gas in massive debris discs is secondary or primordial \citep[e.g.][]{kral-et-al-2019, cataldi-et-al-2023}, constrain the parent planetesimal composition \citep[e.g.][]{kral-et-al-2016}, and distinguish between different gas-production mechanisms \citep[e.g.][]{schneiderman-et-al-2021, bonsor-et-al-2023} by comparing the relative abundances and spatial distributions of carbon-, oxygen-, nitrogen- and hydrogen-bearing species across many systems \citep[e.g.][]{klusmeyer-et-al-2021, smirnov-pinchukov-et-al-2022}. Molecular excitation, isotopologues, and resolved kinematics together constrain the gas temperature, density and velocity field, showing directly how secondary gas evolves once released from planetesimals \citep[e.g.][]{higuchi-et-al-2020}.%, marino-et-al-2026hd121617}.

%The gas distribution also provides a new way of detecting planets during the late stages of planetary system evolution. 
Planets embedded in or interacting with a gaseous debris disc can carve gaps and perturb the otherwise Keplerian velocity field \citep[kinks, e.g.][]{pinte-et-al-2019}, revealing planets too faint for direct imaging, and at separations that are difficult to probe with other exoplanet techniques. Recent modelling shows that planetary gaps may be observable in both CO and C I gas emissions \citep{bergez-casalou-kral-2024}. ALMA2040 will extend this technique to substantially fainter systems and smaller orbital separations, allowing the census of planets inferred from gas structures to extend to older systems and to planetary orbits closer to their host stars.% than can be efficiently probed through the dust distribution \citep[e.g.][]{moor-et-al-2025}. 
%\vspace{-3mm}

%\begin{figure}[ht]
%    \centering
%    \includegraphics[width=1.\linewidth]{co_mass_inc_ALMA2040.png}
%    \vspace{-6mm}
%    \caption{CO mass Vs inclination showing known debris discs. The green (blue) area shows current capability to detect radial gas gaps (kinks), and hence planets in systems with gas. The yellow area shows that this new planet detection technique will further improve thanks to ALMA2040, leading eventually to detections of older and closer-in planets than direct imaging can do.} 
%    \label{fig:gapsplanets}
%\end{figure}

Finally, molecular gas in the terrestrial-planet region probes how volatiles, particularly water, are delivered to planets. Icy planetesimals in young exo-Asteroid belts release water vapour once the primordial disc disperses, forming a secondary gas disc that viscously spreads inward and outward and can deliver substantial water to terrestrial planets without requiring impacts \citep{kral-et-al-2024}. \citet{huet-et-al-2026} predict such discs in a broad range of belts, observable for tens of Myr. ALMA2040 could test whether this mechanism is generic by detecting water in discs down to $10^{-8}$ M$_\oplus$ at 100 pc, extending the accessible parameter space by $\sim$1 order of magnitude. With CO and other volatile tracers, this would link planetesimal reservoirs to planetary volatile inventories and the final assembly of rocky planets \citep{kral-et-al-2020,huet-et-al-2025}.

\vspace{-4mm}

\vspace{-9mm}
\section{Technical Requirements}
\label{sec:TRs}
\vspace{-2mm}
%Summarize the technical requirements here (e.g. angular resolution, spectral resolution, band, bandwidth, sensitivity).

%The main requirement is a much higher sensitivity. The continuum flux, e.g. at 0.9~mm of debris discs, ranges greatly due to their varied distances and dust levels. From large resolving mm-surveys like REASONS (Matr\`a et al. 2025) and ARKS (Marino et al. in press), we know that belts around hotter type stars show surface brightnesses from $10^{-2}$ to $10$~mJy/arcsec$^{2}$. To resolve structures in these discs sufficiently, as shown in Fig.~\ref{fig:debris_disc_sim} and to detect and resolve fainter discs around the cool M dwarfs, we need a sensitivity of 1$\mu$Jy/beam in one hour of observing time. To detect 12CO in terrestrial regions, we need a sensitivity of 0.1 mJy/beam in a 1 km/s channel. Additionally, we need a wider bandwidth to detect more molecules to get a better idea of the gas composition in the inner planetary regions.

A transformative leap in sensitivity is the highest priority: all four science cases are limited mainly by sensitivity.
% how faint a signal can be to still be detectable, which is the main constraint. 
Table~\ref{tab:tech_requirements} lists the capabilities required for each case. All increases are quoted relative to current ALMA (7.5~GHz bandwidth in band~7).

\begin{table}[h]
    %\centering
    \small
    \begin{tabular}{p{0.28\textwidth} p{0.45\textwidth} p{0.18\textwidth}}
    \hline
         \textbf{Capability} & \textbf{Requirement} & \textbf{Science case} \\
    \hline
         Continuum sensitivity & 0.3~$\mu$Jy/beam in 1h at 0.9~mm & §2.1, §2.2, §2.3\\
         Angular resolution & 1~au at 120~pc at 0.9~mm (2$\times$ ALMA) & §2.1, §2.3\\
         Short spacings & MRS 10 -- 20$''$ (60 -- 120~au at 6~pc) & §2.2, §2.3\\
         Line sensitivity & 0.6 -- 3~mm (10$\times$ ALMA) & §2.4\\
         Spectral resolution & 0.1~km/s & §2.4\\
         Bandwidth / multi-band & 512~GHz + multi-band & §2.1, §2.2, §2.3, §2.4\\
    \hline
    \end{tabular}
    \vspace{-1mm}
    \caption{ALMA2040 capabilities required by each science case.}
    \label{tab:tech_requirements}
\end{table}

\vspace{-4mm}
Continuum sensitivity is the key for all three dust-based cases (Sec.~\ref{sec:terrestrial_planet_formation} -- \ref{sec:structures}). %(Sec~\ref{sec:terrestrial_planet_formation}, \ref{sec:m_dwarfs}, and \ref{sec:structures}). 
%It reduces a survey of giant impact candidates from more than 6 yrs with current ALMA to 35~h (Sec~\ref{sec:terrestrial_planet_formation}). For M~dwarfs, it extends the 1~h reach for faint belts from 21 to $\sim$190~pc and for bright belts from $\sim$130~pc to $\sim$1.3~kpc (Sec~\ref{sec:m_dwarfs}). It also brings ten times more belts within reach of high resolution imaging (Sec~\ref{sec:structures}). 
Angular resolution matters most for giant impacts: at the median distance of 120~pc, ALMA2040 reaches $\sim$8~mas (1~au) at 0.9~mm, resolving the terrestrial zones.
For the gas (Sec~\ref{sec:gas}), line sensitivity and spectral resolution must resolve line kinematics and planet signatures. Multi-band observations support all science cases, either by delivering %multiple dust grain-size observations with these 
spectral slopes and grain size distributions (Sec~\ref{sec:terrestrial_planet_formation}, \ref{sec:m_dwarfs}, and \ref{sec:structures}) or %by observing 
gas species and isotopologues (Sec~\ref{sec:gas}). Simultaneous multi-band observing and a wider bandwidth also increase the continuum sensitivity.

\vspace{-5mm}
\section{Synergies with current and future facilities}
\label{sect:synergies}
\vspace{-1mm}
These science cases complement several current and planned facilities, none of which can replace a (sub-)mm interferometer. Only at (sub-)mm wavelengths can we detect the large grains that stay close to their parent planetesimals (small grains are displaced by radiation pressure and stellar winds), and both small and large grain populations are needed to understand grain dynamics \citep[e.g.][]{olofsson-et-al-2022}. The James Webb Space Telescope (JWST) resolves the distribution and composition of small dust grains in the near and mid-IR and has already identified new discs beyond ALMA's reach that are well suited as future ALMA2040 targets \citep{crotts-et-al-2026}.  
\smallskip

\noindent \textbf{\emph{PRIMA} (far-IR).} The Probe far-Infrared Mission for Astrophysics is an accepted NASA far-IR space telescope \citep{glenn-et-al-2025} that will close the current gap in far-IR coverage. It can only resolve the closest debris discs, but its photometry and spectroscopy complement ALMA2040 across all four science cases. 
For giant impacts (Sec.\ \ref{sec:terrestrial_planet_formation}), the far-IR constrains the dust size distribution, and therefore the properties of the colliding bodies \citep{marshall-et-al-2025}.
Its sensitivity makes it a strong complement in the search for M~dwarf belts \citep[Sec.~\ref{sec:m_dwarfs};][]{luppe-et-al-2025}, and its surveys will greatly increase the number of known debris discs \citep{stapelfeldt-et-al-2025,wilner-et-al-2025}, from which targets can be selected for sub-mm follow-up of large grains and planet-disc interaction signatures (Sec~\ref{sec:structures}).
%Finally, PRIMA will provide crucial information to study the gaseous component of debris discs (Sect.\ \ref{sec:gas}). First of all, 
For the gas (Sec.~\ref{sec:gas}), PRIMA can measure the water and CO$_2$ ice content of the grains, which gets released when icy grains collide \citep{chen-et-al-2025}. %. Since gas is produced from the ,destruction of these icy dust grains, constraints on the icy grain composition are crucial to understand debris disc gas. 
It also observes neutral oxygen (O\,I) and ionised carbon (C\,II), the photodissociation products of CO and H$_2$O, complementing ALMA2040 observations of CO, C\,I and H$_2$O
%in terms of studying the origin and physics of debris disc gas 
\citep{cataldi-et-al-2025}.
\smallskip

%\vspace{0.25mm}
\noindent \textbf{\emph{HWO} (optical/near-IR).} The Habitable Worlds Observatory is a concept mission recommended by the Decadal Survey 2020 \citep{Astro20} and recently adopted by NASA. HWO's main mission is to directly image habitable Earth-like planets. Its large collecting area and powerful optical and near-IR coronagraph make it highly complementary to ALMA2040 in the debris disc field.    %, allowing for a better characterisation of the terrestrial regions of planetary systems. 
%While sub-mm observations are necessary to observe large mm-sized grains, the large amounts of 
The micron-sized dust produced in collisional cascades and giant collisions is invisible to ALMA2040. It is needed for a full picture of the dust distribution, and %To fully characterise the dust distribution in debris discs, access to high-spatial-resolution observations at other wavelengths will be fundamental. 
%It will be particularly important for cases where ALMA2040 detects collision-like events through dust and gas clumps \citep[e.g.][]{dent-et-al-2014, roumeliotis-et-al-2026}, as 
HWO will provide that counterpart at high angular resolution.~%enabling a better study of the drag effects. 
This matters most where ALMA2040 detects collision events as dust and gas clumps \citep[e.g.][]{dent-et-al-2014, roumeliotis-et-al-2026}, where comparing small and large grains shows how stellar forces spread the debris. %and the drag effects that shape them.
%Despite not being required for the main mission goal, HWO is expected to carry three other instruments onboard that are yet to be determined. Some of the proposals 
%HWO instruments remain to be selected: proposals include high-resolution spectroscopy \citep[e.g. POLLUX;][]{POLLUX26}, %which again will complement ALMA2040 by enabling dust composition; and observations in the UV, providing extra information about gas species behaviour, similar to science currently produced by the Hubble Space Telescope but with a much larger spatial resolution and sensitivity \citep{Brennan2024}.
%which would provide complementary information on dust composition, while UV capabilities would extend the gas diagnostics, currently provided by \emph{Hubble}, at much larger spatial resolution and sensitivity \citep{Brennan2024}.

\smallskip

\noindent \textbf{Planet detection: ELT and \emph{Gaia}.} The planets that ALMA2040 infers from belt substructures (Sec.~\ref{sec:structures}) can be tested independently. Gaia astrometry \citep{holl-et-al-2023} yields dynamical masses and orbits for giant planets sculpting outer belts, while high-contrast ELT imaging \citep[e.g. METIS;][]{bowens-et-al-2021} can image and characterise them. Pairing ALMA2040 belts with these planets will dynamically weigh the planets through planet-belt interactions, calibrate planet-atmosphere models, reveal their imprints on the belts, and constrain debris disc masses, one of the field's fundamental unknowns \citep[e.g.][]{friebe-et-al-2022, pearce-et-al-2025}.
Gaia's parallaxes, luminosities and kinematic ages provide stellar parameters on which belt temperatures, ages and collisional evolution models depend. The ELT will also help characterise giant impact systems, from IR to mm SED slopes to identify breaks in the grain-size distribution \citep{roumeliotis-et-al-2026}, complementing ALMA 2040's analyses of gas structure, composition, and kinematics. 
\vspace{-6mm}

\section*{Acknowledgements}
\vspace{-4mm}
This work was developed as part of the broader ALMA2040 community effort. We acknowledge the Lorentz Center and NOVA for supporting the community workshops and coordination activities that helped initiate and shape the ALMA2040 White Paper process. PL, ZR and LM  acknowledge funding by the European Union through the E-BEANS ERC project (grant number 100117693), and PL also acknowledges funding by the Irish Research Council (IRC) under grant number IRCLA-2022-3788. SM acknowledges funding by the Royal Society through a Royal Society University Research Fellowship (URF-R1-221669) and the European Union through the FEED ERC project (grant number 101162711). JBL acknowledges the Smithsonian Institute for funding via a CfA Prize Fellowship. AK acknowledges support by the NKFIH NKKP grant ADVANCED 149943. AAS is supported by the Heising-Simons Foundation through a 51 Pegasi b Fellowship.\\

\vspace{-4mm}
%{\footnotesize
\bibliographystyle{apj_mod}
{\bibliography{late-stage_refs}}
%}
\end{document}